# Bandgap properties of low index contrast aperiodically ordered photonic quasicrystals

Gianluigi Zito[1], T. Priya Rose[2], Emiliano Di Gennaro[1], Antonello Andreone[2], Enrico Santamato[1] and Giancarlo Abbate[2]

[1]CNISM and Department of Physics, University of Naples *Federico II*, Napoli, Italy
[2]CNR-INFM *Coherentia* and Department of Physics, University of Naples *Federico II*, Napoli, Italy
Corresponding author: *gianluigi.zito@na.infn.it*

**ABSTRACT:** *We numerically analyze, using Finite Difference Time Domain simulations, the bandgap properties of photonic quasicrystals with a low index contrast. We compared 8-, 10- and 12-fold symmetry aperiodically ordered lattices with different spatial tiling. Our results show that tiling design, more than symmetry, determines the transmission properties of these structures.*

## 1. INTRODUCTION

Quasicrystals are structures exhibiting long-range aperiodic order and rotational symmetry [1, 2]. A quasi-lattice can fill up the entire space to form an infinite structure with forbidden rotational symmetry for periodic crystals. The unit cells overlap each other in a complex manner repeating along the symmetry axis [1]. Although lacking an overall absolute translational symmetry, this local overlapping periodicity acts as a superimposed Bragg grating and gives rise to multiple forbidden photonic band-gap (PBG) frequencies. Photonic QuasiCrystals (PhQCs) may exhibit PBG [3, 4] that are more isotropic than in conventional photonic crystals, permitting the existence of forbidden frequency ranges even in materials possessing very low dielectric contrast. The photonic quasi-lattices are artificial dielectric inhomogeneous media, where scattering centres are usually located in the vertices of the tiles [5]. The quasiperiodic order gives rise to many non equivalent defect sites, increasing the flexibility of these materials for different photonic applications and leading to interesting properties for confining and guiding electromagnetic energy [6, 7]. Superlensing effect, directive emission and mode confinement [8-11] are examples of the quasicrystal features that have recently attracted tremendous interest because of their potential impact in optical components.

Typically, a complete PBG in photonic crystals can be achieved by employing materials with a refractive index contrast substantially higher than zero. Recent studies on quasicrystals with 8-fold (octagonal point group) [3], 10-fold (decagonal) and 12-fold (dodecagonal) [8] rotational symmetries have shown that, in general, most of the PhQCs have wide complete bandgaps even for low threshold values of the refractive index contrast. This characteristic entails the possibility of using a flexible and versatile low cost technology combined with holographic or lithographic fabrication techniques able to realize both large-area and high quality structures in novel soft matter-based and silicon-based devices.

Recently, holographic techniques based on interferential schemes have been widely employed to fabricate PhQCs with high rotational symmetries [12-17]. The light distribution obtained from the interference of two or many coherent light beams is transferred to a photosensitive medium producing the desired quasiperiodic dielectric modulation by single or multiple exposure process. The scattering centres (typically, rods in a binary pattern) of the quasi-lattice correspond to the maxima positions of the light distribution resulting from the *N*-beam interference. The rotational symmetry is determined by the number of beams, whereas the particular tiling pattern of the structure depends on the light intensity, time exposure and on the delicate balance among direction, polarization, amplitude and phase of the beams [15]. By controlling the relative phase delay among the interfering beams, for instance, different geometries in the tiling of the dielectric medium are achievable [16].

Instead, by using direct writing techniques like the electron-beam or single laser beam lithography [18] the spatial tiling of the quasi-lattice can be determined point-by-point. In these cases, usually, the quasiperiodic tiling is previously determined by geometric rules or inflation algorithms. In the case of the octagonal Ammann-Beenker tiling [19, 20] the unit cells consist of "squares" and "rhombuses" of equal side length. Analogously, the Penrose aperiodic tiling may be realized with several approaches. We consider the Penrose tiling consisting of "fat and thin rhombuses", which shows a 10-fold rotational symmetry. Both resulting quasicrystals have 5-fold rotational and mirror symmetry corresponding to a decagonal (10-fold) point group symmetry. Hence, we will refer to the Penrose "rhombus" pattern as decagonal. With regard to the PhQC of 12-fold rotational symmetry, we consider a dodecagonal tiling consisting of "squares" and "triangles" given by Stampfli inflation rule [11].

Although the rotational symmetry of the structure can be of the same order, the quasiperiodic lattices generated by using either interferential methods or geometric rules can exhibit a completely different geometry.

In this work, we numerically analyze using Finite Difference Time Domain (FDTD) simulations the transmittance spectra of 8-, 10- and 12-fold quasicrystal structures. For each fixed rotational symmetry two different quasiperiodic tilings are considered and compared.

## 2. QUASICRYSTAL DESIGN AND SIMULATIONS

Many different types of quasiperiodic structures have been proposed and studied in literature. In this article we compare two different tiling patterns of two-dimensional (2D) symmetric quasi-lattices with 8-, 10-, 12-fold point groups. For each symmetry, the two geometries were determined a) using a geometric or inflation rule, and b) assuming a multiple-beam interference process. Let us call geometric patterns the formers ones and label it with (A), and interferential patterns the latters ones, indicated by (B).

Holographic lithography permits to record large-area photonic quasicrystals in many kinds of photosensitive hard and soft materials, hence it represents an important fabrication technique largely employed to realize high quality quasiperiodic structures. The writing pattern of light is usually obtained as multiple-beam interference and, consequently, the resulting spatial pattern of the dielectric modulation is different from the quasicrystal patterns achievable from inflation tiling. In a typical experimental situation of $N$-beam interference [12, 16], the irradiance profile $I(\mathbf{r})$ achievable according to the relation

$$I(r) = \sum_{m=1}^{N} \sum_{n=1}^{N} A_m A_n^* \exp[i(k_m - k_n) \cdot r + i(\varphi_m - \varphi_n)] \qquad (1)$$

gives the quasiperiodic spatial distributions of the intensity maxima. Here $A_m$, $\mathbf{k}_m$, $\varphi_m$, are the amplitudes, the wave vectors and the initial phases of the interfering beams,
respectively. The wave vectors $\mathbf{k}_m$ of the beams, for $m=(1, \ldots, N)$, are oriented, according to the relation

$$k_m = \frac{2\pi n}{\lambda} \left( \sin\left(\frac{2\pi n}{N}\right) \sin\theta, \cos\left(\frac{2\pi n}{N}\right) \sin\theta, \cos\theta \right) \qquad (2)$$

at an angle $\theta$ with respect to the longitudinal direction along $z$-axis and are equally distributed along the transverse ($x$, $y$)-plane. $n$ is the average refractive index of the medium and $\lambda$ is the common wavelength of the beams. Usually, the beams are supposed to have the same linear polarization. Their number $N$ is related to the rotational symmetry of the quasicrystal [17]. By changing the phase delay between the beams, different patterns having the same symmetry can be realized [16, 21]. The filling factor, defined as the ratio between the area of $n_H$ (high refractive index) regions and total area, depends on the threshold level of the photosensitive material and the exposure time and intensity. Typically, the maxima positions of the light pattern correspond to the high dielectric regions, that can be usually approximated with a structure of homogenous dielectric rods embedded in a host medium of different index.

Fig. 1 shows the detail of the six different structures that were analyzed, in which the number of dielectric elements was held fixed to 900. The 8-fold geometric structure octagonal (A) simulated and shown in Fig. 1-a consists of dielectric rods (high index $n_H$ medium) in air (low index $n_L$ medium) located at the vertices of the Ammann-Beenker tiling of space, that is the positions of the cylinders of radius $r$ are coincident with the vertices of "squares" and "45° rhombuses" with sides of equal length. Recently, it was theoretically found that octagonal (A) pattern presents a complete PBG with a very low threshold value for the refractive index contrast, suggesting the possibility to realize optoelectronic devices based on this tiling pattern in silica or even in soft materials like polymer [20]. In Fig. 1-b, the 8-fold interferential pattern is shown, calculated for $N=8$ from Eq. (1) and (2). The structure was obtained imposing a particular phase delay between the beams, that is the phases were periodically shifted by $\pi/2$ so that $\varphi_1=\varphi_5=0$, $\varphi_2=\varphi_4=\varphi_6=\varphi_8=\pi/2$ and $\varphi_3=\varphi_7=\pi$. The pattern was obtained by positioning circular rods of radius $r$ (top view in Fig. 1) in the maxima of the continuous irradiance distribution $I(\mathbf{r})$ resulting from Eq. (1) [22]. The quasicrystal lattice is therefore determined without calculating a particular tiling of the plane. Nevertheless, the spatial tiling can be easily derived after having generated the structure. The patterns depicted in Fig. 1-a and 1-b are clearly different. Although they possess the same rotational symmetry, the tiling originating the pattern provides a different assembly of the dielectric elements. The particular interferential pattern of Fig. 1-b, octagonal(B), was studied in a preliminary work [22].

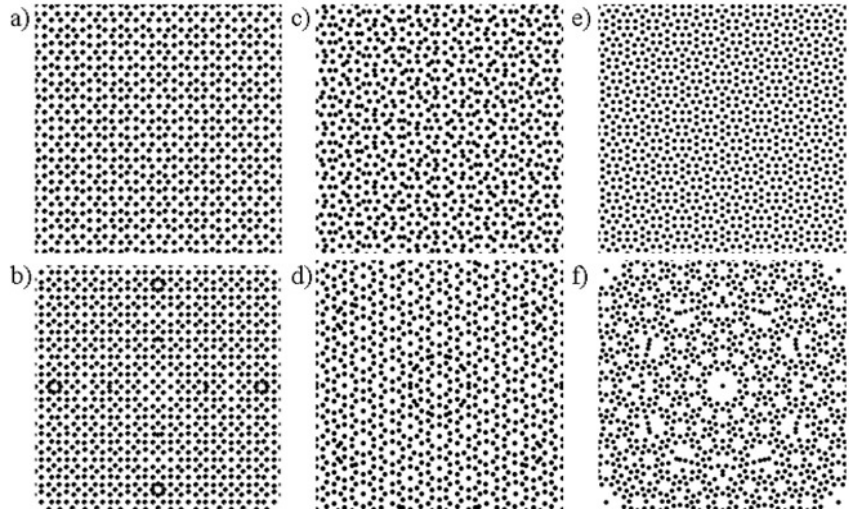

**Figure 1** (a)-(b) Octagonal patterns of rods (top view) with geometric square-rhombus tiling (Ammann-Beenker) and interferential tiling, respectively; (c)-(d) decagonal patterns with geometric rhombic tiling (Penrose) and interferential tiling, respectively; (e)-(f) dodecagonal geometric triangle-square tiling (Stampfli inflation rule) and interferential tiling, respectively.

The patterns shown in Fig. 1-c and 1-d are 10-fold quasicrystals obtained from the geometric Penrose tiling with rhombic cells [23] of equal side length (decagonal (A)), and supposing a 10-beam interference (decagonal (B)), respectively. In Fig. 1-e and 1-f, the dodecagonal (A) and (B) patterns obtained from inflation algorithm and 12-beam interference, respectively, are shown. In particular, dodecagonal (A) is a 12-fold quasicrystal formed with triangle-square tiling by a recursive algorithm and scaled up by an inflation factor, that was found to induce a photonic bandgap even in host medium with low threshold index like glass [4]. Multiple-beam patterns of both 10-fold and 12-fold symmetry have been calculated from Eq. (1) and (2) supposing a typical holographic process of fabrication in which the beams have the same linear polarization and equal optical phase, that is $\varphi_1=...=\varphi_8=...=\varphi_{12}$.

Due to the non-geometric building of the patterns depicted in Fig. 1-b, 1-d and 1-f, we found useful to define a new parameter, that is the average distance *a* between neighbouring rods along the *x*-direction, as characteristic length of the patterns. This parameter was used to design the geometric patterns too. The filling factor is related to the ratio *r/a*, where *r* is the rod radius. Such parameter was held fixed to *r/a*=0.18 in each simulation to permit the comparison between the structures analyzed here.

## 3. TRANSMISSION PROPERTIES OF QUASICRYSTAL STRUCTURES

As aforementioned, two different tiling patterns, obtained from geometric tiling and interference-based method, respectively, were analyzed for octagonal, decagonal and dodecagonal symmetry.

The two-dimensional FDTD method with Perfectly Matched Layer (PML) boundary conditions (along *x*- and *y*-direction) was used in all simulations [20]. The FDTD technique was employed to obtain transmission information, through the (*x*, *y*)-plane, as a function of propagation direction and wavelength, for both TE (electric field $E_z$ perpendicular to the lattice plane) and TM (magnetic field $H_z$ perpendicular to the lattice plane) polarization. In each numerical simulation a Gaussian time-pulse excitation was launched from outside the structure, or inside it for comparison. Several detectors (time monitors) were placed in specific positions allowing to store the field components in correspondence of a particular propagation direction of the light source [24]. Their positions were chosen to cover the angular range related to the 8-, 10- and 12-fold rotational symmetry with an angular separation from 5° to 15° depending on the structure under study. The Fourier Transform of the time-dependent signal collected by the detector and normalized with respect to the incident light provided the frequency response of the structure with a high resolution. The corresponding transmission spectra had a wavelength range between 0.1 and 6.0μm with a resolution of $\delta=5.0\times10^{-4}$μm, whereas the discretization grid provided a minimum of 100 grid points per free space wavelength. The transmission coefficient was calculated as a function of the refractive index difference $\Delta n$ for each propagation direction of the time-pulse source. The data collected from the detectors placed at different positions and angular orientations presented the same overall shape and band-gap properties, demonstrating that the structures are almost isotropic with respect to the propagation direction. The transmission spectra reported here were then obtained by taking the data collected from a single detector by fixing the incidence direction.

In Fig. 2, the transmittance spectra related to the octagonal (A) (left panel) and octagonal (B) (right panel) patterns, obtained for increasing values of the refractive index difference $\Delta n$, in particular 0.4 (a)-(b), 0.6 (c)-(d) and 0.8 (e)-(f), are shown as a function of normalized wavelength $\lambda/a$, for both TE (black curve) and TM (red curve) polarization. In the case of the octagonal (A) tiling, the bandgap starts to appear even at $\Delta n$=0.4 (see Fig. 2-a), but for TE polarization only. As the dielectric contrast increases, the attenuation of the transmission signal in the bandgap region enhances from ~13dB to ~30dB (see Fig. 2-a and 2-e) with an increase of the width to midgap ratio $\Delta\lambda/\lambda_m$ from 2.4% at $\Delta n$=0.4 to 14.4% at $\Delta n$=0.8. The attenuation was estimated by averaging the values obtained in correspondence of the spectral gap. The shift in the position of the bandgap, evident from the variation of the normalized midgap wavelength $\lambda_m/a$ from 1.22 to 1.38, is reasonable since increasing values of $\Delta n$ correspond to an increase in the average refractive index of the quasicrystal medium. In the case of the octagonal (B) tiling, the bandgap starts to appear at $\Delta n$=0.6, centered at $\lambda_m/a$=1.79, once again for TE polarization only (Fig. 2-d). In Fig. 2-f, two bandgaps are instead visible corresponding to $\Delta n$=0.8. The first one is centered at the normalized wavelength $\lambda_m/a$=1.61 and the other one at $\lambda_m/a$=1.93. The first bandgap has $\Delta\lambda/\lambda_m$~5% and presents a substructure of peaks that might be related to the existence of localized states. Very interesting is the presence of strong transmittance attenuations, of the order of ~50dB and ~30dB, found in the gap regions at $\lambda_m/a$=1.61 and $\lambda_m/a$=1.93, respectively.

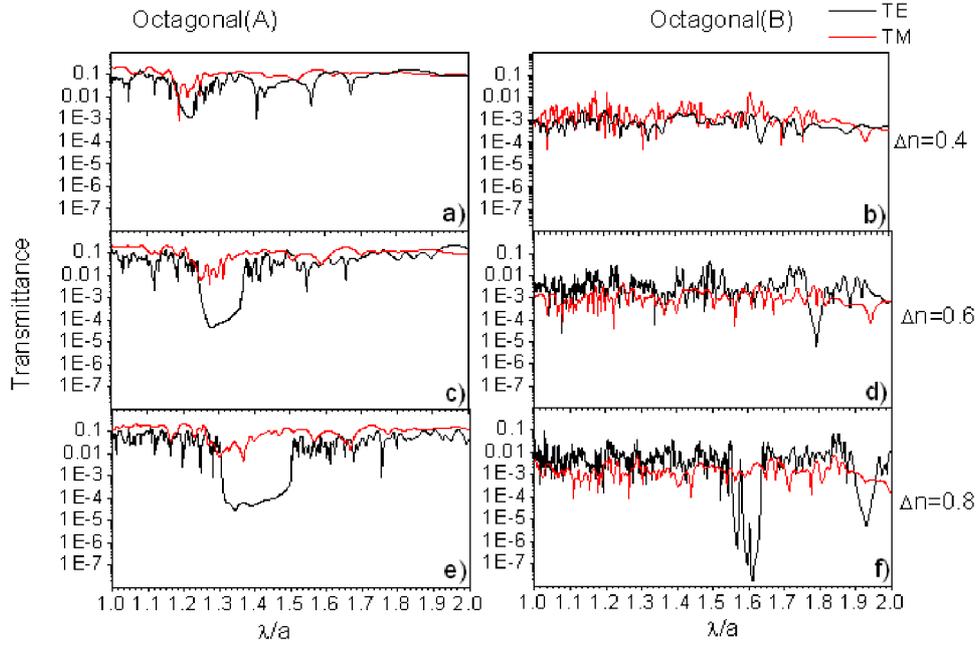

**Figure 2** Transmittance spectra calculated for 8-fold symmetry structures with geometric tiling, octagonal (A) (left panel), and interferential tiling, octagonal (B) (right panel), for TE (black curves) and TM (red curves) polarization, and increasing values of the refractive index difference: $\Delta n$=0.4 (a)-(b), $\Delta n$=0.6 (c)-(d), $\Delta n$=0.8 (e)-(f).

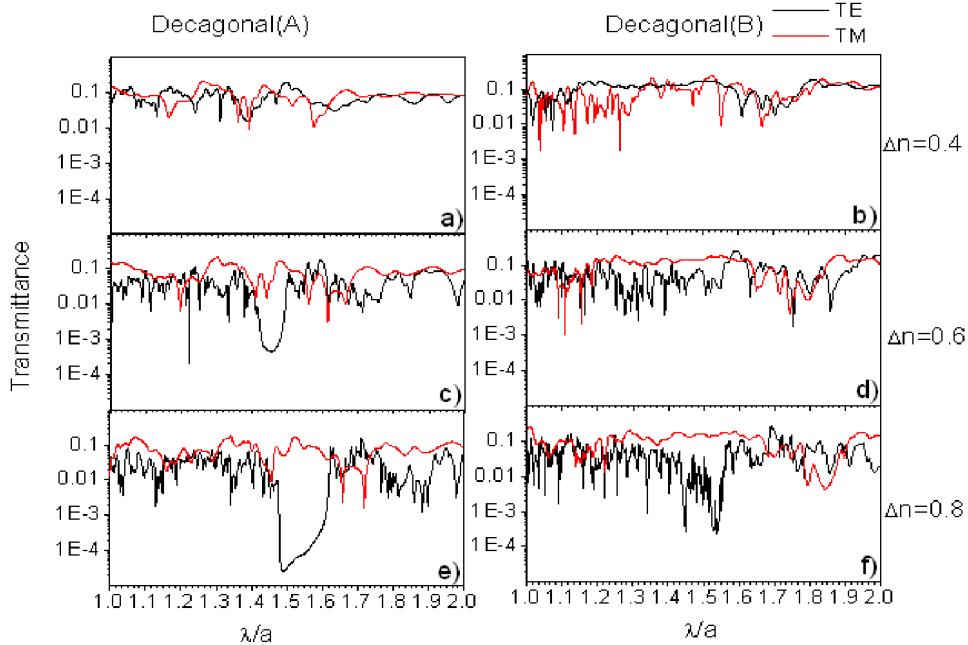

**Figure 3** Transmittance spectra calculated for 10-fold symmetry structures with geometric tiling, decagonal (A) (left panel), and interferential tiling, decagonal (B) (right panel), for TE (black curves) and TM (red curves) polarization, and increasing values of the refractive index difference: $\Delta n$=0.4 (a)-(b), $\Delta n$=0.6 (c)-(d), $\Delta n$=0.8 (e)-(f).

The same procedure was applied to derive the bandgap properties of the decagonal quasicrystalline structures. The corresponding TE (black line) and TM (red line) transmittance spectra for both decagonal (A) (left panel) and decagonal (B) (right panel) patterns, obtained for Δ$n$=0.4 (a)-(b), 0.6 (c)-(d) and 0.8 (e)-(f) as a function of normalized wavelength λ/$a$, are presented in Fig. 3. For TE polarization, the decagonal (A) pattern shows a tiny bandgap at a refractive index difference Δ$n$=0.4 with a ~5dB attenuation (see Fig. 3-a), that enhances to ~20dB as the index difference increases to Δ$n$=0.6, with Δλ/λ$_m$=3.7% at λ$_m$/$a$=1.45 (see Fig. 3-c). The signal attenuation further increases to ~30dB with a normalized width Δλ/λ$_m$=9% at λ$_m$/$a$=1.5 for Δ$n$=0.8, as reported in Fig. 3-e. No clear PBG was found for TM polarization, at least for the values of dielectric contrast examined here. As evident from the right panel of Fig. 3, the decagonal (B) pattern, on the other hand, does not show any clear bandgap for both polarizations at Δ$n$=0.4 and 0.6 (see Figs. 3-b and 3-d). For Δ$n$=0.8, a feature with ~20dB attenuation is visible in the transmission spectrum for TE polarization at λ$_m$/$a$~1.5 in a narrow range in which many other dips, associable to defect states, are present. This "bandgap" decreases even further for TM polarization (~10dB attenuation at λ$_m$/$a$~1.8), as shown in Fig. 3-f.

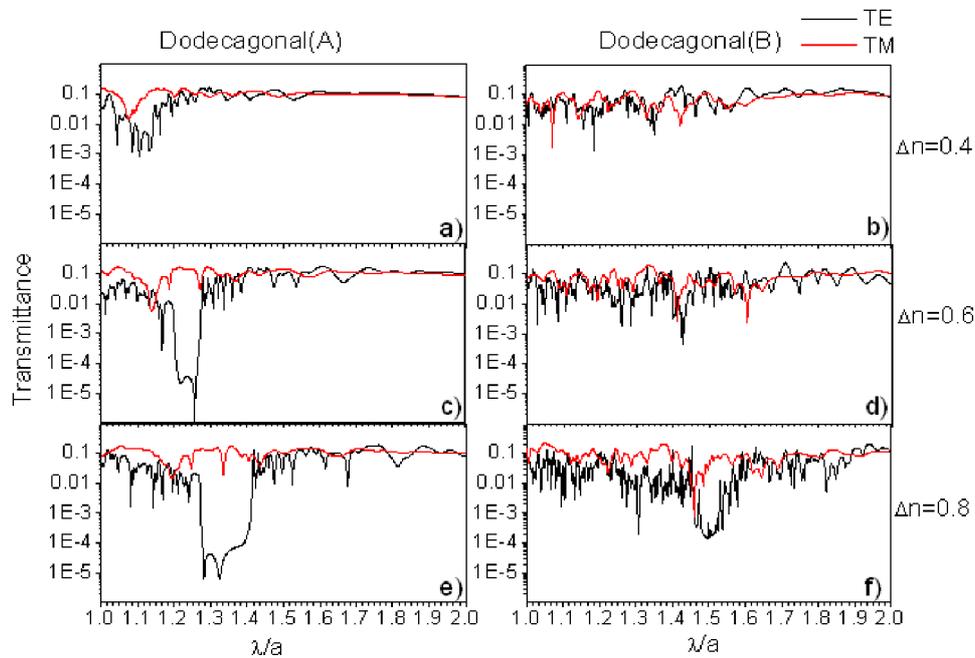

**Figure 4** Transmittance spectra calculated for 12-fold symmetry structures with geometric tiling, dodecagonal (A) (left panel), and interferential tiling, dodecagonal (B) (right panel), for TE (black curves) and TM (red curves) polarization, and increasing values of the refractive index difference: Δ$n$=0.4 (a)-(b), Δ$n$=0.6 (c)-(d), Δ$n$=0.8 (e)-(f).

The transmittance spectra associated to the 12-fold quasicrystals, dodecagonal (A) and (B), were calculated and reported in the left and right panels of Fig. 4 respectively, analogously to the other structures previously analyzed. For the dodecagonal (A) tiling pattern, the bandgap starts to open up at Δ$n$=0.4 (Fig. 4-a) for TE polarization, becoming wider at Δ$n$=0.6 with a normalized width Δλ/λ$_m$= 5.7% at λ$_m$/$a$=1.24 and ~25dB of attenuation in the transmission coefficient (see Fig. 4-c). The bandgap is shifted to λ$_m$/$a$=1.34 with an increased width of Δλ/λ$_m$= 10.3% and an attenuation of ~30dB at Δ$n$=0.8 (see Fig. 4-e). The interferential pattern (dodecagonal (B)) shows only a very narrow bandgap, for TE polarization and index difference as low as Δ$n$=0.6 in correspondence of the normalized midgap wavelength λ$_m$/$a$=1.43. The bandgap width increases to Δλ/λ$_m$=4.5%, slightly shifted to λ$_m$/$a$=1.50, at Δ$n$=0.6 and with an attenuation < 20dB, although few localized states are visible within this spectral region, as reported in Fig. 4-f. Also in this case, no remarkable photonic bandgap is observed in both (A) and (B) patterns for TM polarization of the time-pulse excitation.

From these simulations, it appears clear that the patterns under study – obtained from geometric algorithms or resulting from a multiple-beam holographic process – show remarkable differences not only with regard to the PBG properties but also in relation to the existence of the bandgap. The differences arisen from the transmittance spectra clearly point out the importance of the tiling geometry in determining the photonic bandgap properties of the quasicrystal, independently of the dielectric contrast and the rotational symmetry of the quasi-lattices studied in this work.

## 4. CONCLUSIONS

In this work, we analyzed the formation and development of the photonic bandgap in 2D 8-, 10- and 12-fold symmetry quasicrystalline lattices of low dielectric contrast. In particular, two different quasiperiodic patterns were considered and compared for each fixed order of symmetry. Our numerical simulations of the transmittance spectra, based on the finite difference method, prove that the different tiling geometry, more than the rotational symmetry, dramatically affects the existence and behaviour of the photonic bandgap in low dielectric contrast structures. If wide PBGs are to be obtained for small $\Delta n$, hence, very accurate control of the fabrication parameters is mandatory. The patterns that we examined here were chosen to highlight the substantial and remarkable PBG differences that arise between a quasicrystalline geometric tiling and a similar quasicrystal achievable from holographic fabrication techniques. Holographic lithography represents, in fact, an efficient and feasible fabrication method able to provide large-area photonic quasicrystals of high quality both in soft and hard materials [12-16]. In this direction, we have recently developed a single-beam holographic technique able to provide the desired tiling patterns in order to realize high efficiency PBG structures for photonic applications [21, 22].